\documentstyle[aps,twocolumn,epsf,psfig]{revtex}

\begin{document}

\twocolumn[\hsize\textwidth\columnwidth\hsize\csname @twocolumnfalse\endcsname



\title{Subgap   structures   in  the  current-voltage  characteristic   of 
the intrinsic Josephson effect due to phonons} 

\author{  Ch.~Helm,  Ch.~Preis, F.~Forsthofer  and J.~Keller,  \\ 
Institut  f\"ur Theoretische  Physik,  Universit\"at  Regensburg, 
D-93040 Regensburg,  Germany \\ K.  Schlenga, R.  Kleiner, and P. 
M\"uller,   \\   Physikalisches   Institut   III,   Universit\"at 
Erlangen-N\"urnberg, D-91058 Erlangen, Germany} 

\date{\today}

\maketitle

\begin{abstract}  

A modified RSJ-model for the coupling of intrinsic Josephson oscillations 
and c-axis phonons in the high-$T_c$ superconductors
Tl$_2$Ba$_2$Ca$_2$Cu$_3$O$_{10+\delta}$ and 
Bi$_2$Sr$_2$CaCu$_2$O$_{8+\delta}$ is developed.
This provides a very good explanation for recently reported subgap structures
in the $I$-$V$-characteristic of the c-axis transport. 
It  turns out that the voltages of these structures coincide with the 
eigenfrequencies of longitudinal optical phonons, providing a new 
measurement technique for this quantity. The significantly enhanced 
microwave emission at the subgap structures in both the GHz and THz region
is discussed.

\end{abstract}   

]




The intrinsic Josephson effect between the CuO$_2$ (multi)-layers in 
highly anisotropic cuprate superconductors has been demonstrated in a 
characteristic series of branches in the current-voltage characteristic in  
c-direction \cite{kleiner}.
Each  branch  corresponds  to a well  defined  number  of Josephson 
junctions in the resistive state.  
Recently the observation of subgap   structures   in  the 
$I$-$V$-characteristic  of  
Tl$_2$Ba$_2$Ca$_2$Cu$_3$O$_{10+\delta}$        (TBCCO)        and 
Bi$_2$Sr$_2$CaCu$_2$O$_{8+\delta}$      (BSCCO)     has     been 
reported   \cite{yurgens,schlenga}.   The   structures   in   the 
different branches can be traced back to the $I$-$V$-characteristic 
of one single Josephson  junction in the resistive  state.  These 
structures  seem  to be an intrinsic  effect, as they  have been 
observed in junctions of different geometries.
Furthermore 
the characteristic  voltages of the subgap structures are  completely 
independent  of temperature  and external  magnetic fields, which 
rules out any relation to the superconducting gap, vortex flow or 
the thermal excitation of quasiparticles. 

In this paper we will  show that these  subgap  structures  can very
well  be  explained  by the  coupling  of c-axis  phonons  to the 
Josephson field oscillations  in a single resistive junction.
Thereby it turns out that  the maxima
in the $I-V$-characteristic  correspond  to the  eigenfrequencies  of 
longitudinal lattice vibrations. A more detailed discussion can be found in 
\cite{wir}.

In the RSJ model the total  current density  through  a junction 
\begin{equation}  
I/F  = j_c  \sin  \gamma  + j_{qp}(E) +  {\dot  D}  \;  . \label{rsj}  
\end{equation} 
is the sum of the Josephson current, the quasiparticle  current and 
the displacement  current density. The (gauge 
invariant)  phase difference  $\gamma$ is related to the electric 
field  $E$  in the  barrier  of thickness  $b$  by the  Josephson 
relation 
\begin{equation}
\hbar \dot \gamma = 2eEb \; . \label{josephson}
\end{equation} 
In the simplest case the quasiparticle  current  density  is 
described  by an  ohmic law $j_{qp}= \sigma E$.  Here we will use a
more general form $j_{qp}(E)= j_{qp}(\bar E) + \sigma (E(t) - \bar E)$
with $\sigma = j_{qp}^{\prime} (  {\bar E} ) $ by 
expanding the field $E(t)$ around the time averaged field $\bar E$.  
The displacement  current density 
${\dot D}$  contains  the polarization  $P$ of the barrier  medium,  $D= 
\epsilon_0  E + P =\epsilon_0  \epsilon  E$.  In the case of high 
frequency   Josephson   oscillations   in  the  range  of  phonon 
frequencies it is important to keep the full frequency dependence 
of $\epsilon  (\omega)$  or to treat  the polarization  $P$ as an 
additional dynamical variable. 

Here we assume that the polarization $P$ is due to a 
c-axis displacement $z$ 
of ions  with  charge  $q$, mass $M$ and density $n$ in the
non superconducting  barrier
between  the copper  oxide (multi)-layers.
For  the  motion  of  the  ions  we  assume  a simple  oscillator 
\begin{equation}
\ddot   z  +  \omega^2_0   z  +  \rho  \dot   z  =  {q \over   M}   E ,
\label{oscillator}
\end{equation} 
which is driven by the electric field $E$ in the barrier. 
In this model the contribution  of 
the oscillator to the dielectric function is given by 
\begin{equation}
\epsilon ( \omega )  =  \epsilon_\infty 
+  \frac{S\omega_0^2}{\omega_0^2 -\omega^2 - i \rho  \omega}
\;   \label{dk}
\end{equation}
where  $S=  n   q^2/(\epsilon_0   M  \omega_0^2)$   is  the 
oscillator   strength   of  the  phonon and  $\epsilon_\infty$  includes
the other  contributions to the polarizability  of the barrier.  
In  this  form it is straightforward to generalize the calculation to
several phonon branches.

The system is characterized by the transversal optical
phonon frequency $\omega_0$, 
the Josephson  plasma  freqency   
$\omega_J^2 =   2ebj_c/(\hbar\epsilon_0 \epsilon_{\infty}) $
  and a characteristic 
frequency  $\omega_c=(2e/\hbar)V_c$, $V_c$  being  the voltage 
 at the critical  current $j_c$.   In our case $\omega_J  < 
\omega_0  < \omega_c$  and  the  McCumber  parameter  $\beta_c  = 
\omega_c^2/\omega_J^2$ is large.  

Using equ. (\ref{josephson}) in (\ref{rsj}) and (\ref{oscillator}) 
we obtain a coupled set of 
differential   equations  for  the  phase $\gamma$  and  the  polarization 
amplitude $z$, which can be solved numerically. Then the time average 
of the phase velocity $\langle \dot\gamma \rangle    = 
(2e/\hbar) V $ determines  the dc voltage $V$ 
shown  in the $I-V$ characteristic. 

The most important features of the model can also be understood analytically: 
Due to the large McCumber parameter $\beta_c$
both the phase and the polarization   
\begin{eqnarray}  
\gamma (t)   &  \approx  & \gamma_0  + \omega  t +
\gamma_1  \sin\omega  t, \\ 
z (t)  & \approx &  z_0  + z_1  \cos  (\omega  t + \varphi) \; .  
\end{eqnarray} 
oscillate  primarily  with the basic frequency $\omega = (2e/\hbar)V$.
It is also well known  that in this case $\gamma_1 \ll 1$ and 
$\sin (\gamma (t) ) $ can be expanded in orders of $\gamma_1$ keeping only the 
leading Fourier components:
\begin{equation}
\sin \gamma(t) \approx \sin \gamma_0 \cos \omega t + \cos \gamma_0 
\sin \omega t - {1\over 2} \gamma_1 \sin \gamma_0.
\end{equation}

Since  the equation  of motion  for the phonon  is linear,  
we can relate the phonon amplitude to the driving force $\dot \gamma$ by 
the phonon response function of equ.
(\ref{dk}) at the frequency $\omega$. Thus 
we obtain for the coefficents  of
$\sin\omega t$, $\cos \omega t$ and the dc current $i= I/I_c$:
\begin{eqnarray}  
0 &=& \cos\gamma_0 +
{ \omega^2\over \omega_J^2 } \epsilon_1 \gamma_1 \; , \\
0 &=& \sin \gamma_0 + { \omega^2\over \omega_J^2 } \left( \epsilon_2 + {\sigma
\over  \epsilon_0\omega } \right)   \gamma_1  \; , \\  
i &=& i_{qp}(V) - {1\over 2} \gamma_1 \sin \gamma_0 \; , 
\end{eqnarray}  
where $\epsilon_1, \epsilon_2$ are the real and imaginary part of
the phonon dielectric function. From these three equations we finally 
obtain for the normalized dc-current 
\begin{eqnarray}
i (V)  &=& i_{qp}(V)  +    {1\over  2}  {\omega_J^2 \over  \omega^2} 
\frac{ \epsilon_2 + \frac{\sigma}{\omega \epsilon_0}  }{
\epsilon_1^2 + 
 \left( \epsilon_2 + \frac{ \sigma }{ \omega \epsilon_0}  \right)^2 } \\
&=&  i_{qp}(V)  -  {1\over  2}  {\omega_J^2 \over  \omega^2}  {\rm Im} 
\left( 1 \over \tilde \epsilon(\omega )   \right) 
\end{eqnarray} 
with  $\tilde   \epsilon   =  \epsilon_1   +  i  (\epsilon_2   + 
\sigma/(\omega\epsilon_0))$ and $\omega = (2e/\hbar)V$.

This result clearly  shows that maxima in the $I-V$ characteristic 
occur  near  the  zeros  of  the  phonon dielectric function $\epsilon_1$ 
exactly 
at the eigenfrequencies of longitudinal optical phonons with wavevector 
${\vec q} = 0 $.  Here the polarization 
and  the  Josephson  field  oscillations  are  compensating  each 
other,  and the energy  transfer  to the phonon is maximal.
In contrast to this, the Josephson oscillation amplitude 
$\gamma_1$ has a minimum at the transversal optical eigenfrequency  
$\omega_0$.

In figures ({\ref{TBCCOkennlinie}) and (\ref{BSCCOkennlinie})  this
theory is compared with  experimental   results 
\cite{schlenga} by choosing appropriate   values   for   the 
McCumber parameter $\beta_c$,   the  phonon  frequencies $\omega_{0,i}$,
their oscillator strength $S_i$ and damping $\rho_i$, and the  quasi-particle 
characteristic.   We see that the essential features are very well 
reproduced.   We also find  an excellent  agreement  between  the 
numerical  integration  and  the analytical  results.
Note that both in experiment and theory hysteretic regions
 appear near the phonon resonances.

In TBCCO the characteristic voltage is given by $V_c = 27.1 \; 
{\rm mV} $ and the
two maxima correspond to LO-phonon frequencies 
$\nu_{\rm LO, 1}=3.65 \; {\rm THz}$ and $\nu_{\rm LO, 2}=4.70 \;  {\rm THz}$.
The corresponding values for BSCCO are: $V_c = 21.8 \; {\rm mV}$, 
  $\nu_{\rm LO, 1}= 2.96 \; {\rm THz}$,
$\nu_{\rm LO, 2} = 3.90 \; {\rm THz}$ and $\nu_{\rm LO, 3} = 5.71 \; 
{\rm THz}$.
These values are compatible  with phonon resonances  observed  in 
the   infrared   reflectivity   at  3.7   THz  and  4.3  THz   of 
polycristalline samples of TBCCO \cite{phononTBCCO}. 
Unfortunately  for  both  TBCCO  and BSCCO  no reliable  infrared 
experiments for c-axis phonons in single crystals are available 
\cite{phononTBCCO,phononBSCCO}.
Lattice dynamical calculations \cite{kulkarni}  show phonon modes 
with large oscillator strength, 
where TlO layers are oscillating 
against  CuO layers  at somewhat  higher  frequencies.
The dispersion of these LO-phonons in c-direction is small, which causes 
a small, but finite damping $\rho$ in the model described above.

\begin{figure}
\leavevmode
\vspace{-1.1cm}
\begin{center}
\epsfxsize=0.5\textwidth
\epsfbox{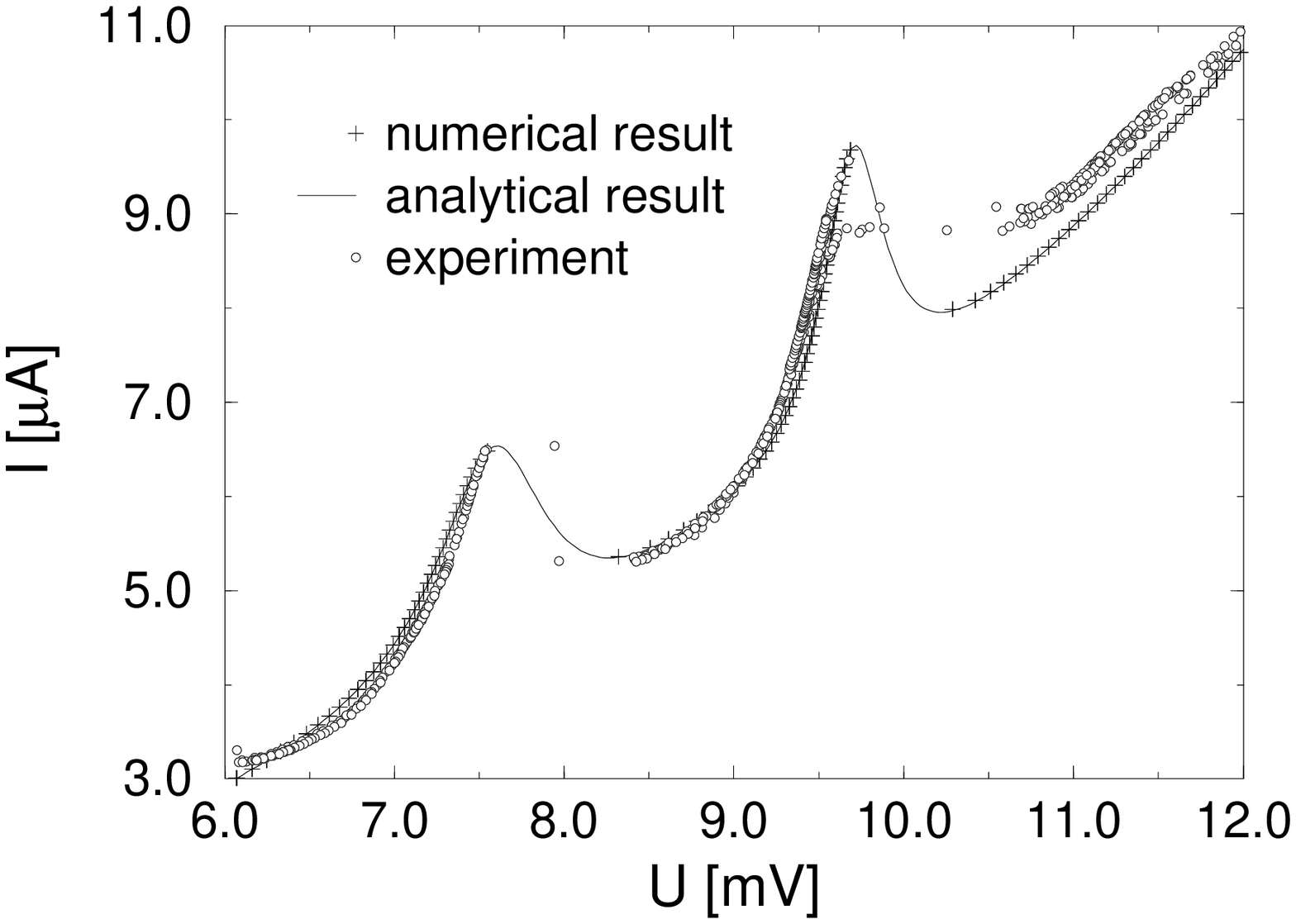}
\caption{\label{TBCCOkennlinie}\mbox{Experimental, analytical and numerical}  
        $I$-$V$-cha\-rac\-ter\-istic of TBCCO for $\beta_c=375$, 
        $\omega_{0,1} = 3.28 \; _{\rm THz}$, 
        $S_1= 2.87$, $\rho_1=0.2 \; {\rm THz}$, $\omega_{0,2}= 4.45 \; 
        {\rm THz}$,
        $S_2=0.68 $, 
        $\rho_2 = 0.4 \; {\rm THz} $, $\epsilon_{\infty} = 8.45$.} 
\end{center}
\end{figure}     
\begin{figure}      
\leavevmode 
\vspace{-2.3cm}      
\begin{center}      
\epsfxsize=0.5\textwidth 
\epsfbox{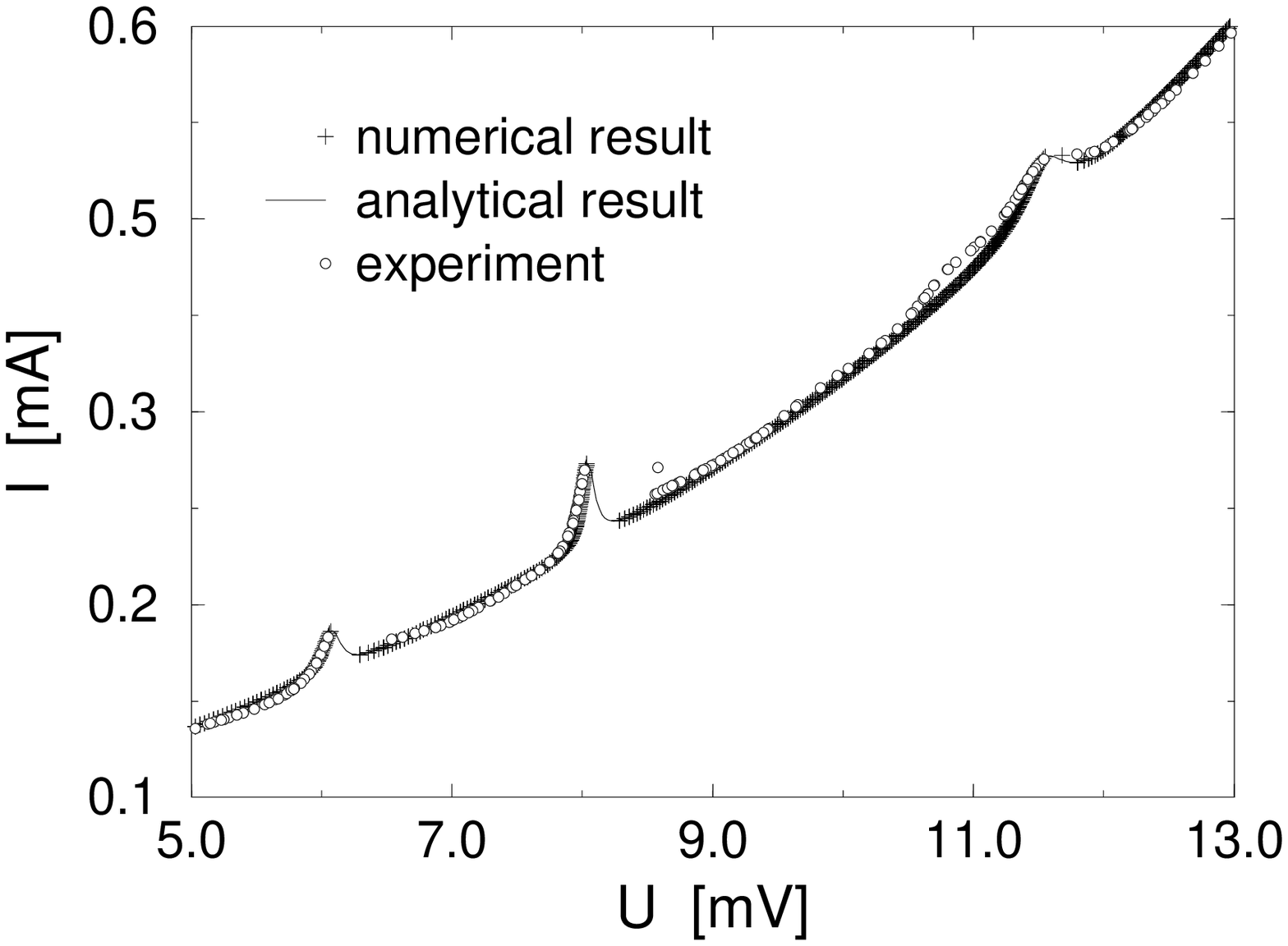} 
\caption{\label{BSCCOkennlinie}\mbox{Experimental, analytical and 
numerical}    $I$-$V$-cha\-rac\-ter\-istic     of    BSCCO    for 
$\beta_c=800$, $\omega_{0,1} = 2.80 \; {\rm THz}$, 
$S_1 = 1.35 $, $\rho_1=0.08 \; 
{\rm THz} $, 
$\omega_{0,2} = 3.62 \; {\rm THz}$, $S_2 = 1.5$, $\rho_2=0.05 \; {\rm THz}$, 
$\omega_{0,3} = 4.91 \; {\rm THz}$, $S_3 = 1.65$, $\rho_3=0.2 \; {\rm THz} $, 
$\epsilon_\infty= 7.5$. } 
\end{center} 
\vspace{-0.4cm} 
\end{figure} 

A similar calculation can be used to determine the microwave power 
radiated by Josephson oscillations: the emission is enhanced  
by about one order of magnitude near the maxima of the subgap structures, 
suggesting these voltages for an experimental attempt to demonstrate the 
microwave emission of the intrinsic Josephson oscillations directly. 
Also the observed enhancement of the microwave emission in the 
GHz region, if the junction is biased near the subgap structures, 
can be explained: the vanishing differential conductivity 
near the maxima of the $I-V$-curve leads to an amplification of 
voltage fluctuations caused by thermal fluctuations of the current.

Further theoretical investigations suggest that also the mixing of two 
external frequencies (in the THz region) down to the difference frequency 
(GHz) is much more efficient, if the external frequencies 
are near the maxima of the subgap structures.  
 
The modifications of the intrinsic Josephson oscillations through  the
dielectric constant of the phonons  is in complete formal 
analogy with the coupling of a macroscopic Josephson junction to an external 
electromagnetic system with a resonance, i.e. a cavity. 
In such systems hysteretic structures 
in the $I$-$V$-characteristic, which correspond to the subgap structures 
considered  in this publication, have been reported repeatedly \cite{likharev}.

To summarize,
the recently discovered subgap structures in the $I$-$V$-characteristic 
in the intrinsic Josephson-effect in high-$T_c$-superconductors are explained 
by a coupling of Josephson oscillations 
to optical c-axis phonons within a modified RSJ-model. 
Apart from perfect reproduction of the experimental data for both 
BSCCO and TBCCO  new  insights in the physical interpretation
of the peak structures are obtained.
The peak position is identified with 
the eigenfrequency of the longitudinal optical phonon, which  provides 
a new possibility of a direct measurement of this quantity, which is 
normally hard to determine in optical experiments. Significantly enhanced 
microwave emission near the maxima of the subgap structure 
in the GHz region is explained and theoretically predicted in the THz region.

{\it Acknowledgment:} This work has been supported by a grant 
of the Bayerische Forschungsstiftung  within the research program 
FORSUPRA and  by the Studienstiftung des Deutschen Volkes (C.H.).

\end{document}